\documentclass[twocolumn,english,aps,prl,reprint,floatfix,notitlepage,footinbib,preprintnumbers,superscriptaddress]{revtex4-1}
\pdfoutput=1
\usepackage{lmodern}

\usepackage[T1]{fontenc}
\usepackage[latin9]{inputenc}
\usepackage{geometry}
\usepackage{subfigure,lmodern, amsmath,amssymb, graphicx, pifont, adjustbox, bm, xcolor}
\usepackage{amsfonts}
\usepackage{geometry}
\usepackage{comment}
\usepackage{mathtools}
\usepackage{float}
\usepackage{slashed}
\usepackage{ragged2e}
\usepackage{array}

\makeatletter\g@addto@macro\bfseries{\boldmath}\makeatother

\usepackage{stackengine}
\usepackage{esint}
\usepackage[unicode=true,pdfusetitle,
 bookmarks=true,bookmarksnumbered=false,bookmarksopen=false,
 breaklinks=false,pdfborder={0 0 1},backref=false,colorlinks=true]
 {hyperref}
\hypersetup{pdftitle={UV-Complete Gravity Amplitudes and the Triple Product},
 pdfauthor={Yu-tin Huang and Grant N. Remmen},
 citecolor=black,linkcolor=black,urlcolor=black}

\newcommand{\appendixref}[1]{\hyperref[#1]{appendix~\ref{#1}}}
\def\equationautorefname~#1\null{eq.\,(#1)\null}
\usepackage{breakurl}
\usepackage{breakurl}
\usepackage[hang,flushmargin]{footmisc} 
\allowdisplaybreaks
\makeatletter

\usepackage{etoolbox}
\apptocmd{\thebibliography}{\justifying\setlength{\leftskip}{7.4mm}}{}{} 
 
 \usepackage{relsize}
\usepackage{babel}

\makeatletter
\def\simgt{\mathrel{\lower2.5pt\vbox{\lineskip=0pt\baselineskip=0pt
           \hbox{$>$}\hbox{$\sim$}}}}
\def\simlt{\mathrel{\lower2.5pt\vbox{\lineskip=0pt\baselineskip=0pt
           \hbox{$<$}\hbox{$\sim$}}}}
\makeatother

\usepackage{changepage}

\newcommand{\eq}{\begin{equation}}
\newcommand{\eqe}{\end{equation}}
\newcommand{\eqa}{\begin{eqnarray}}
\newcommand{\eqae}{\end{eqnarray}}
\newcommand{\be}{\begin{equation}}
\newcommand{\ee}{\end{equation}}
\newcommand{\bea}{\begin{eqnarray}}
\newcommand{\eea}{\end{eqnarray}}

\newcommand{\Eq}[1]{Eq.~(\ref{#1})}

\newcolumntype{P}[1]{>{\centering\arraybackslash}p{#1}}

\begin{document}

\pagestyle{plain}

\title{UV-Complete Gravity Amplitudes and the Triple Product}

\author{Yu-tin Huang}
\affiliation{Department of Physics and Astronomy, National Taiwan University, Taipei 10617, Taiwan}
\thanks{e-mail: \\ \url{yutinyt@gmail.com}, \url{remmen@kitp.ucsb.edu}}
\affiliation{Physics Division, National Center for Theoretical Sciences, Taipei 10617, Taiwan}
\author{Grant N. Remmen}
\affiliation{Kavli Institute for Theoretical Physics, University of California, Santa Barbara, CA 93106, USA}
\thanks{e-mail: \\ \url{yutinyt@gmail.com}, \url{remmen@kitp.ucsb.edu}}
\affiliation{Department of Physics, University of California, Santa Barbara, CA 93106, USA}

\begin{abstract}
\noindent 
We construct an infinite class of new ultraviolet-complete four-graviton scattering amplitudes that reduce to Einstein gravity at low energies, vanish at high energies, are meromorphic, and exhibit a triple-product structure ${\cal A}(s){\cal A}(t){\cal A}(u)$. The spectrum invariantly exhibits accumulation points in the form of infinite towers of states on each mass pole, whose residue can be positively expanded on tree-level exchanges of irreducible representations of the Lorentz group. 
\end{abstract}
\maketitle

\noindent{\bf Introduction.}---At energies below the Planck scale, general relativity can be treated as an effective field theory (EFT), with Feynman rules computable in perturbative quantum gravity. However, at high energies, perturbative unitarity breaks down. This is easily visible from the tree-level four-graviton scattering amplitude, 
\be
{\cal M}_{\rm grav} = \frac{\kappa^2 \mathbb{R}^4}{stu}, \label{eq:Mgravintro}
\ee
where $\kappa^2 = 8\pi G$ is the gravitational coupling, $\mathbb{R}^4$ is a kinematic invariant of polarization tensors that can be written as the contraction of four linearized Riemann tensors~\footnote{Specifically, 
\begin{equation*}
\begin{aligned}
\mathbb{R}^4 =&\, 32 (R_1)^{\mu\nu\rho\sigma}(R_2)_{\mu\;\;\rho}^{\;\;\alpha\;\;\beta}(R_3)_{\alpha\;\;\nu}^{\;\;\gamma\;\;\delta}(R_4)_{\beta\gamma\sigma\delta} \\&- 8 (R_1)^{\mu\nu\rho\sigma}(R_2)_{\mu\nu}^{\;\;\;\;\alpha\beta}(R_3)_{\rho\alpha}^{\;\;\;\;\gamma\delta}(R_4)_{\sigma\beta\gamma\delta},
\end{aligned}
\end{equation*}
plus all other permutations of the labels $1,2,3,4$, where $(R_i)^{\mu\nu\rho\sigma}$ is the linearized Riemann tensor, $(p_i^\nu p_i^\sigma\epsilon_i^{\mu\rho} {-} p_i^\nu p_i^\rho \epsilon_i^{\mu\sigma} {-} p_i^\mu p_i^\sigma \epsilon_i^{\nu\rho} {+} p_i^\mu p_i^\rho \epsilon_i^{\nu\sigma})/2$. Alternatively, for scattering scalars minimally coupled to gravity, we simply replace $\mathbb{R}^4$ with $(s^4 + t^4 + u^4)/2$.}, and in terms of incoming momenta, the Mandelstam variables are $s\,{=}\,{-}(p_1{+}p_2)^2$, $t\,{=}\,{-}(p_2{+}p_3)^2$, and $u\,{=}\,{-s}\,{-}\,t$ for a mostly-plus metric. Since ${\cal M}_{\rm grav} \propto E^2/m_{\rm Pl}^2$, the amplitude diverges with characteristic scale $m_{\rm Pl}$, leading to a breakdown of perturbative unitarity beyond the Planck scale directly analogous to the failure of perturbative unitarity in $WW$ scattering at the weak scale in the absence of the Higgs.

A UV-complete gravitational theory is expected to provide a unitarized four-graviton amplitude; therefore, consistency constraints on the latter presents an avenue to \textit{bootstrap} the theory of quantum gravity. Constraints such as unitarity and causality are often encoded in the analytic behavior of the scattering amplitude. Let us consider a perturbative completion, in which case the graviton loops are suppressed, so that the amplitude is analytic, but for poles, at low energies. Indeed, working with tree-level completions has led to tremendous insight into the UV spectrum stemming from causality constraints~\cite{Camanho:2014apa}, as well as the EFT beyond the leading Einstein-Hilbert action from the combined requirements of causality, unitarity, and crossing~\cite{Bern:2021ppb, Caron-Huot:2021enk, Chiang:2021ziz, Caron-Huot:2022ugt, Chiang:2022jep, Cheung:2016wjt}.  
Moreover, as shown via $\hbar$ counting in Ref.~\cite{Cheung:2016wjt}, any theory that perturbatively unitarizes Einstein-Hilbert graviton scattering must do so at tree level.

Another obvious reason for considering tree-level amplitudes is that perturbative string theories with $1/\sqrt{\alpha'}\,{\ll}\, m_{\rm Pl}$, along with their deformations~\cite{Arkani-Hamed:2020blm}, provide explicit examples of tree-level completions of gravity. Consider the celebrated four-graviton amplitude for the type-II superstring,
\eq
{\cal M}_{\text{type{-}II}}=-\frac{\kappa^2 \mathbb{R}^4\Gamma({-}\tfrac{\alpha'}{4}s)\Gamma({-}\tfrac{\alpha'}{4} t)\Gamma({-}\tfrac{\alpha'}{4} u)}{\Gamma(1\,{+}\, \tfrac{\alpha'}{4} s)\Gamma(1\,{+}\, \tfrac{\alpha'}{4}t)\Gamma(1\,{+}\, \tfrac{\alpha'}{4}u)},\label{eq:typeII}
\eqe
which reproduces Eq.~\eqref{eq:Mgravintro} at low energies. An interesting feature of this solution is that it is a triple product,
\eq
{\cal M}_{\text{type{-}II}}=\kappa^2 \mathbb{R}^4 {\cal A}(s){\cal A}(t){\cal A}(u),\label{eq:triple}
\eqe
where ${\cal A}(s)\,{=}\,{-}\Gamma({-}\tfrac{\alpha'}{4}s)/\Gamma(1\,{+}\,\tfrac{\alpha'}{4} s)$. Note that ${\cal A}(s)$ has a simple pole with residue $({-}1)^{n}/(n!)^2$ at nonnegative integer $n$ and vanishes as $s\rightarrow{+}\infty$. The four-${\rm U}(1)$ amplitude of the heterotic string and four-tachyon amplitude of the bosonic string can also written as triple products. Taking this fact as inspiration, we ask if there exist other UV completions of the graviton amplitude that admit the form~\eqref{eq:triple}. As it turns out, there are an infinite number of such amplitudes if one permits accumulation points, by which we mean, following the terminology of Refs.~\cite{Bern:2021ppb,Caron-Huot:2016icg}, a theory with an infinite number of states at a given mass.

Accumulation-point amplitudes appear to play a special role in low-energy EFT bounds. Indeed, for a massless scalar, it was shown in Ref.~\cite{Caron-Huot:2020cmc} that a triple-product form with a single pole, ${\cal A}(s)=1/(m^2{-}s)$,  plays a special role in spanning the space of consistent EFTs (see Ref.~\cite{Figueroa:2022onw} for the role of Coon's amplitude~\cite{Coon:1969yw,Fairlie:1994ad} in such a context). 
This amplitude contains an infinite tower of higher-spin states of mass $m$, since the pole at $s\,{=}\,m^2$ in ${\cal A}(s){\cal A}(t){\cal A}(u)$ becomes an infinite series in $\cos \theta$ for scattering angle $\theta$.
The existence of such accumulation-point amplitudes was also conjectured to be important in spanning the gravitational EFT~\cite{Chiang:2022jep}. In this work, we aim to construct such amplitudes.
In our ansatz, we are modeling a tree-level completion and thus consider a spectrum with distinct, isolated poles, i.e., meromorphic amplitudes, in contrast to Coon's generalization of the Veneziano amplitude, which involves an infinite accumulating sequence of distinct poles and thus fails to be analytic.
Motivated by the behavior of Eq.~\eqref{eq:typeII}, we will require our ${\cal A}(s)$ to contain only simple poles and vanish as $s\rightarrow +\infty$.

While we do not identify a model---e.g., a Hamiltonian or Lagrangian description---corresponding to our amplitudes, taking the bootstrap-like approach of directly building the amplitudes themselves first can be useful in understanding the necessary characteristics of a UV completion of gravity. Such an approach has the notable historical precedent of the construction of the Veneziano amplitude~\cite{Veneziano:1968yb}, which preceded its interpretation as a theory of strings~\cite{Susskind:1969ha}.

This paper is structured as follows.
We construct an infinite-parameter generalization of the massive-pole product amplitude for scalar scattering of Ref.~\cite{Caron-Huot:2020cmc} and prove that it is unitary.
Using this as a starting point, we then build UV-soft gravitational amplitudes with a particular triple-product structure.
Computing the partial waves for arbitrary spectra, we confirm that our amplitudes are unitary.
We subsequently consider the large-spin limit and discuss soft theorems, causality, and a potentially related construction in string theory.
\newline

\noindent {\bf Triple-product~amplitudes.}---Consider a function describing the general $s$-channel exchange of an arbitrary tower of states:
\be 
\bar{\cal A}(s)=\sum_{n}\frac{g_{n}^{2}}{-s+m_{n}^{2}},\label{eq:A}
\ee
where the $g_{n}$ are arbitrary real couplings and the $m_{n}^{2}>0$ describe an arbitrary spectrum of distinct masses.
From $\bar{\cal A}(s)$, let us define a four-point amplitude for massless scalar scattering,
\be 
\overline{\cal M}=\bar{\cal A}(s)\bar{\cal A}(t)\bar{\cal A}(u).\label{eq:A1}
\ee
By construction, $\overline{\cal M}$ is Bose-symmetric. 

The residues of $\overline{\cal M}$ have a positive expansion on the Gegenbauer polyonomials, as required by unitarity, in arbitrary spacetime dimension $D$.
To see this, let us define the residue at $s\,{=}\,m_{n}^{2}$ as ${\cal R}_n =\lim_{s\rightarrow m_{n}^{2}}(-s+m_{n}^{2})\overline{\cal M} =\sum_{i,j} {\cal R}^{ij}_n$, where 
\be 
{\cal R}^{ij}_n=\frac{g_{n}^{2}g_{i}^{2}g_{j}^{2}}{\left(-t+m_{i}^{2}\right)\left(t+m_{n}^{2}+m_{j}^{2}\right)}. \label{eq:Rij}
\ee
Note that the sum in $i,j$ will pick out the symmetric part of ${\cal R}_n^{ij}$. We further define the rescaled mass parameters, 
\be 
M_{i,n}^{2} =1+\frac{2m_{i}^{2}}{m_{n}^{2}} >1,\label{eq:Ms}
\ee 
and write the scattering angle in terms of $\cos\theta\,{=}\,x$ (where $t\,{=}\,m_n^2(x\,{-}\,1)/2$ on the massive pole), in terms of which the symmetrized residue is
\be 
{\cal R}_n^{(ij)} =\frac{4 g_n^2 g_i^2 g_j^2}{m_n^4(M_{i,n}^{2}{+}M_{j,n}^{2})}\frac{M_{i,n}^{2}}{M_{i,n}^{4} - x^{2}} + (i\leftrightarrow j),\label{eq:RR}
\ee
where $T^{(ij)}\,{=}\,(T^{ij} \,{+}\, T^{ji})/2$. Since $M_{i,n}^{2}, M_{j,n}^{2}\,{>}\,1$ and $|x|\,{\leq}\, 1$, we can rewrite the result as a geometric series in $x^2$ resulting in a positive-definite polynomial. 
Since the monomial $x^{2k}$ has a strictly positive expansion in the Gegenbauer polynomials,
\be 
x^{2k}=\sum_{\ell=0}^{k}\frac{(2k)!(\alpha{+}2\ell)\Gamma(\alpha{+}1)}{\alpha2^{2k}(k{-}\ell)!\Gamma(\alpha{+}1{+}k{+}\ell)}C_{2\ell}^{(\alpha)}\!(x),\label{eq:monomial}
\ee
this immediately implies that the residue has a positive partial-wave expansion in any dimension, so $\overline{\cal M}$ in Eq.~\eqref{eq:A1} is indeed unitary.

It will be useful to have the explicit form of the partial-wave expansion, ${\cal R}_n(x){=}\sum_{\ell=0}^{\infty}a^{(n)}_{\ell}C_{\ell}^{(\alpha)}\!(x)$,
where $C_{\ell}^{(\alpha)}$ are the Gegenbauer polynomials and $\alpha=(D-3)/2$. 
Odd-$\ell$ partial waves will vanish since $t\leftrightarrow u$ crossing symmetry implies that ${\cal R}$ is even in $x$.
Let us compute a useful integral for even $\ell$,
\be 
\begin{aligned}
&I_{\ell}^{(\alpha)}\!(v)=\int_{{-}1}^{{+}1}\frac{(1{-}x^{2})^{\alpha{-}\frac{1}{2}}}{v^{2}{-}x^{2}}C_{\ell}^{(\alpha)}\!(x){\rm d}x \\
&\;\;=\frac{\pi\Gamma(\ell{+}2\alpha)\,{}_{2}F_{1}\!\left(\tfrac{\ell+1}{2},\!\tfrac{\ell+2}{2};\ell{+}1{+}\alpha;\tfrac{1}{v^{2}}\right)}{2^{2\alpha+\ell-1}v^{\ell+2}\Gamma(\alpha)\Gamma(\ell{+}1{+}\alpha)}.
\end{aligned}
\ee
Using orthogonality of the Gegenbauer polynomials, $
\int_{{-}1}^{{+}1}C_{\ell}^{(\alpha)}\!(x)C_{m}^{(\alpha)}\!(x)(1{-}x^{2})^{\alpha{-}\frac{1}{2}}{\rm d}x \,{=}\, \delta_{\ell m}/F_{\ell}^{(\alpha)}$,
where $F_{\ell}^{(\alpha)}{=}\{\ell!(\ell{+}\alpha)[\Gamma(\alpha)]^{2}\}/[\pi 2^{1{-}2\alpha}\Gamma(\ell{+}2\alpha)]$ is a normalization, we find the even-spin partial waves,
\be 
a_{\ell}^{(n)}=\frac{8 g_n^2 F_{\ell}^{(\alpha)}}{m_n^4}\sum_{i,j}\frac{g_i^2 g_j^2M_{i,n}^{2}I_{\ell}^{(\alpha)}\!(M_{i,n}^{2})}{M_{i,n}^{2}{+}M_{j,n}^{2}},\label{eq:alR}
\ee
which are all positive since $I_\ell^{(\alpha)}\!(v)\,{>}\,0$.
The amplitude $\overline{\cal M}$ therefore describes the unitary exchange of an infinite tower of even higher-spin states at each of the masses $m_n$.
\newline

\noindent {\bf Gravitational triple product.}---We will now find that there is a very intriguing variation on the triple-product amplitude introduced above that will allow us to accommodate gravity. In particular, consider
\be
{\cal A}(s) = \frac{1}{s} + \sum_{n=1}^\infty \frac{g_n^2}{-s+m_n^2} \label{eq:Agrav}
\ee
for $g_n,m_n$ arbitrary parameters and (nonzero real) masses, respectively~\footnote{The string amplitude in \Eq{eq:typeII} is not of the form in \Eq{eq:Agrav}. Instead, $-\Gamma({-}s)/\Gamma(1{+}s)$ can be written as $\Sigma_{n=0}^\infty f_n(s)/({-}s{+}n)$, where $f_n(s) = [\Pi_{i{=}0}^n ({-}2s{-}1{+}i)]/[n! \Gamma(2{+}2s)]$,
i.e., the $g_n^2$ are replaced in ${\cal M}_{\text{type{-}II}}$ by analytic functions of $s$ of indefinite sign on each pole in ${\cal A}$.}, and where the sum can also be taken to be finite.
As before, we take the sequence of $m_n$ to be free of accumulation points, so that ${\cal A}$ is meromorphic.
Imposing a single requirement on the $g_n$,
\be
\sum_{n=1}^\infty g_n^2 = 1, \label{eq:gconstraint}
\ee
we will find the remarkable result that the triple-product amplitude,
\be 
{\cal M} = \kappa^2 \mathbb{R}^4{\cal A}(s){\cal A}(t){\cal A}(u),\label{eq:MM}
\ee
with ${\cal A}$ as given in \Eq{eq:Agrav}, unitarizes graviton scattering. That is, ${\cal M}$ in Eq.~\eqref{eq:MM} reduces to ${\cal M}_{\rm grav}$ in the IR while satisfying perturbative unitarity in the UV and obeying unitarity with all nonnegative partial waves on each pole.

The coupling constraint~\eqref{eq:gconstraint} enables perturbative unitarity, since for high-energy fixed-angle scattering ${\cal M} \,{\propto} \,(1{-}\sum_n g_n^2)^3 E^2$, while with Eq.~\eqref{eq:gconstraint} this is dramatically softened to ${\cal M}\,{\propto}\, E^{-4}$.
Similarly, in the Regge limit of large $s$ and fixed $t$, ${\cal M} \propto (1{-}\sum_n g_n^2)^2 s^2$ for general couplings, but with Eq.~\eqref{eq:gconstraint} this improves to ${\cal M} \propto s^0$, in both cases satisfying the bound in Ref.~\cite{Chowdhury:2019kaq}.
\newline

\noindent {\bf Partial waves.}---Let us compute the partial waves for our amplitude ${\cal M}$ on its various massless and massive residues, all of which we must prove are positive to guarantee unitarity. Unitarity of string amplitudes was established in Refs.~\cite{Pius:2016jsl,Sen:2016bwe,Sen:2016uzq}.
Recently, Ref.~\cite{Arkani-Hamed:2022gsa} further showed that the $\mathbb{R}^4$ itself prefactor can be written as a positive expansion of exchanges of irreps in $D\,{\leq}\,10$.  Thus, if the residue of the triple product ${\cal A}(s){\cal A}(t){\cal A}(u)$ itself has a positive expansion on the Gegenbauer polynomials, then this tells us that the residue of ${\cal M}$ corresponds to the exchange of tensor products of irreps with positive coefficients.

We note that convergence of the partial-wave expansion will as usual require $D \,{\geq}\, 5$ due to the $t$-channel singularity. (That is, $t\propto \theta^{-2}$ in the forward limit, while the volume measure goes like ${\rm d}\Omega \sim \theta^{D-3} {\rm d}\theta$, leading to a partial-wave divergence in $D\,{=}\,4$~\cite{Giddings:2009gj}.)
As before, $t\leftrightarrow u$ crossing symmetry ensures that only even-$\ell$ partial waves contribute. 

Let us first consider the residue on the massless $s\,{=}\,0$ pole, ${\cal R}_0 = \lim_{s\rightarrow 0} (-s){\cal A}(s){\cal A}(t){\cal A}(u)$, 
\be
\begin{aligned}
{\cal R}_0 &= \bigg(\frac{1}{t}{+}\sum_{i=1}^{\infty}\frac{g_{i}^{2}}{-t\,{+}\,m_{i}^{2}}\bigg)\bigg(\frac{1}{t}{+}\sum_{j=1}^{\infty}\frac{g_{j}^{2}}{{-}t\,{-}\,m_{j}^{2}}\bigg)\\
&=\!\sum_{i,j} g_{i}^{2}g_{j}^{2}\bigg(\frac{1}{t}{+}\frac{1}{{-}t\,{+}\,m_{i}^{2}}\bigg)\!\bigg(\frac{1}{t}{+}\frac{1}{{-}\,t{-}\,m_{j}^{2}}\bigg),
\end{aligned}
\ee
where in the second line we used Eq.~\eqref{eq:gconstraint}.
Rewriting as a series and rearranging, we obtain
\be
{\cal R}_0 = \!\sum_{i,j} \!\frac{g_i^2 g_j^2}{m_{i}^{2}{+}m_{j}^{2}}\sum_{\substack{\ell\geq0 \\{\rm even}}}^{\infty} t^{\ell{-}2}\bigg(\frac{m_{i}^{2}}{m_{j}^{2\ell}}{+}\frac{m_{j}^{2}}{m_{i}^{2\ell}}\bigg) >0.
\ee
The overall $\mathbb{R}^4$ scales like $t^4$, so the $t^{-2}$ term in the residue corresponds to the spin-two Einstein-Hilbert contribution, while the remaining terms form an infinite series with even degree in $t$ and all positive coefficients. If we insist on interpreting these extra terms as the exchange of higher-spin states, the positivity of its coefficient is consistent with unitarity. To see this, we will use $D\,{=}\,4$ for simplicity, even though the partial-wave expansion will ultimately require $D\,{\geq}\,5$. For $(+,+,-,-)$ helicity configuration, the $s$-channel exchange of an even-$\ell$ state has residue given by  
\be 
\begin{aligned}
{\cal M}(1^{+2},2^{+2},P^{+\ell})\times {\cal M}(3^{-2},4^{-2},{-}P^{{-}\ell})\\
=\frac{([1P][P2])^\ell}{[12]^{\ell{-}4}}\frac{(\langle3P\rangle\langle P4\rangle)^\ell}{\langle34\rangle^{\ell{-}4}}=[12]^4\langle 34\rangle^4t^\ell,
\end{aligned}
\ee
where we have used $P\,{=}\,p_3\,{+}\,p_4$ to simplify to the last equality, and the spinor brackets correspond to $\mathbb{R}^4$ in this helicity arrangement. Thus, on the $s\,{=}\,0$ pole sits an infinite number of massless even higher-spin states. 
In our amplitude~\eqref{eq:MM}, we are not considering these exchanged particles as isolated, asymptotic states; we will come back to the issue of external massless higher spins in the discussion.

Let us next consider the residue our triple product on the massive poles. At $s\,{=}\,m_n^2$, the residue ${\cal R}_n \,{=}\, \lim_{s\rightarrow m_n^2}({-}s\,{+}\,m_n^2){\cal A}(s){\cal A}(t){\cal A}(u)$ can be written as
\be
g_{n}^{2}\bigg(\frac{1}{t}{+}\sum_{i=1}^{\infty}\frac{g_{i}^{2}}{m_{i}^{2}{-}t}\!\bigg)\!\!\bigg(\!\frac{-1}{m_{n}^{2}{+}t}{+}\sum_{j=1}^{\infty}\frac{g_{j}^{2}}{t{+}m_{n}^{2}{+}m_{j}^{2}}\!\bigg).
\ee
This residue has three qualitatively different types of terms: those ${\propto}\,g_i^2 g_j^2$ from the product of massive propagators, those ${\propto}\,g_i^2$ or $g_j^2$ alone from the cross-terms involving massive and massless poles, and those independent of $g_i^2$ and $g_j^2$ from the $1/tu$ product of massless propagators.
That is, we can write
\be 
{\cal R}_{n}={\cal R}_{n}^{00}+\sum_{i=1}^{\infty}{\cal R}_{n}^{0i}+\sum_{i=1}^{\infty}\sum_{j=1}^{\infty}{\cal R}_{n}^{ij},
\ee 
where ${\cal R}_{n}^{00}  =-g_n^2/[t(t\,{+}\,m_{n}^{2})]$,
\be 
{\cal R}_{n}^{0i}  =\frac{g_{n}^{2}g_{i}^{2}}{t(t+m_{n}^{2}+m_{i}^{2})}+\frac{g_{n}^{2}g_{i}^{2}}{(t+m_{n}^{2})(t-m_{i}^{2})},
\ee
and ${\cal R}_n^{ij}$ is given in \Eq{eq:Rij}.
From Eq.~\eqref{eq:RR},
\be 
{\cal R}_{n}^{(ij)} {=}\,\frac{4g_{n}^{2}g_{i}^{2}g_{j}^{2}}{m_{n}^{4}(M_{i,n}^{2}{+}M_{j,n}^{2})}\sum_{k=0}^{\infty}\frac{x^{2k}}{M_{i,n}^{4k+2}} {+} (i\leftrightarrow j).\label{eq:Rnijexpand}
\ee
Meanwhile, for the ${\cal R}^{00}_n$ part of the residue, we can expand in a series in $x$ to obtain
\be 
{\cal R}_n^{00}(x) = \frac{4g_n^2}{m_n^4(1-x^2)} =  \frac{4g_n^2}{m_{n}^{4}}\sum_{k=0}^{\infty}x^{2k},\label{eq:R00expand}
\ee
while the cross-term residue for ${\cal R}_n^{0i}$ can be written as
\be 
{\cal R}_n^{0i}= -\frac{8g_n^2 g_i^2}{m_{n}^{4}(1{+}M_{i,n}^{2})}\sum_{k=0}^{\infty}\bigg(1{+}\frac{1}{M_{i,n}^{4k+2}}\bigg)x^{2k}.\label{eq:R0iexpand}
\ee
While Eqs.~\eqref{eq:Rnijexpand} and \eqref{eq:R00expand} are strictly positive, the cross-term~\eqref{eq:R0iexpand} is negative.

Using our requirement on the couplings in Eq.~\eqref{eq:gconstraint}, we assemble the full residue into the form
\be 
{\cal R}_{n}=\frac{4g_{n}^{2}}{m_{n}^{4}}\sum_{k=0}^{\infty}\sum_{i=1}^{\infty}\sum_{j=1}^{\infty}g_{i}^{2}g_{j}^{2}Q_{nijk}x^{2k},
\ee 
where we can rearrange $Q_{nijk}$ into a manifestly positive form (recalling
that $M_{i,n}^{2},M_{j,n}^{2}>1$),
\begin{equation}
Q_{nijk}{=}\frac{(M_{i,n}^{2}\!{-}1)(M_{j,n}^{4k{+}4}\!{-}1)}{M_{j,n}^{4k{+}2}(1{+}M_{j,n}^{2})(M_{i,n}^{2}\!{+}M_{j,n}^{2})} {+} (i{\leftrightarrow} j).\label{eq:Q}
\end{equation}
By Eq.~\eqref{eq:monomial}, Eq.~\eqref{eq:Q} immediately implies that all of the partial waves have nonnegative coefficients.
Explicitly, in the partial-wave expansion ${\cal R}_n(x) \,{=} \sum_{\ell{=}0}^{\infty}a_{\ell}^{(n)}C_{\ell}^{(\alpha)}\!(x)$, we have $a_\ell^{(n)} {=} \,0$ for odd $\ell$, while for even $\ell$, $a_\ell^{(n)} {=}\, 8g_n^2 m_n^{-4}\sum_{i,j} g_{i}^{2}g_{j}^{2} A_\ell^{(n)ij}$,
where 
\be 
\begin{aligned}
A_\ell^{(n)ij} & {=} \frac{F_\ell^{(\alpha)}}{2(1{+}M_{i,n}^2)}\!\bigg[\frac{M_{i,n}^2(1{-}M_{j,n}^2)I_\ell^{(\alpha)}\!(M_{i,n}^2) }{M_{i,n}^2 {+} M_{j,n}^2}
\\& \qquad\qquad\;\;\;\;\;\;\, {+}\frac{M_{i,n}^2 {-} 1}{2}I_\ell^{(\alpha)}\!(1)\bigg] {+} (i{\leftrightarrow} j).\label{eq:AA}
\end{aligned}
\ee
All of the $A_\ell^{(n)ij}$ are positive for arbitrary massive spectra.
As a result, $a_\ell^{(n)} > 0$, and so ${\cal M}$ represents a unitary UV completion of ${\cal M}_{\rm grav}$.
\newline

\begin{figure}[t]
\begin{center}
\includegraphics[width=0.95\columnwidth]{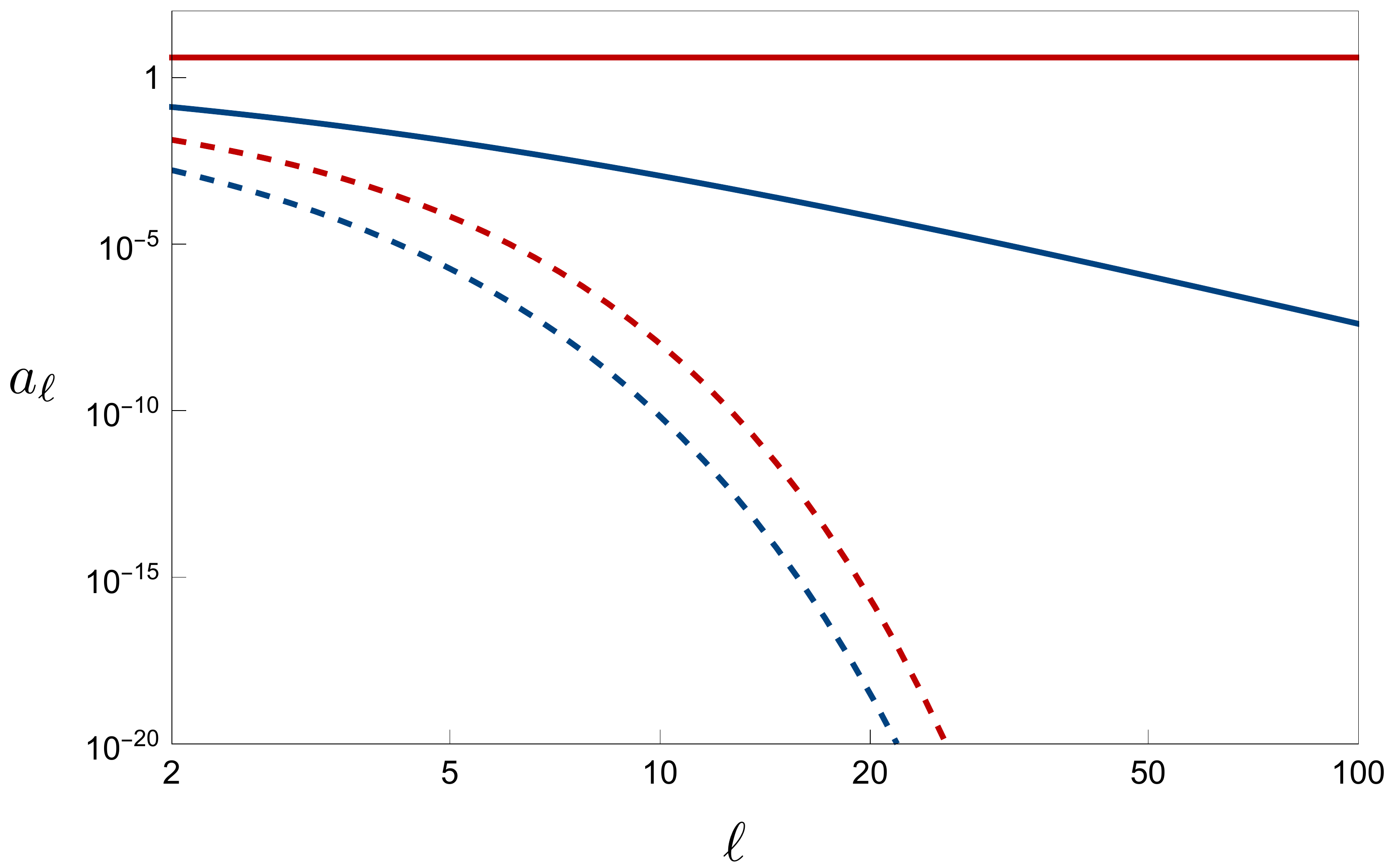}
\end{center}\vspace{-7mm}
\caption{Partial waves for the gravitational (solid) and nongravitational (dashed) amplitudes calculated in text in Eqs.~\eqref{eq:alR} and \eqref{eq:AA}, for example theories with a single massive pole at $m\,{=}\,1$ and coupling $g\,{=}\,1$, for $D\,{=}\,5$ (red) and $D\,{=}\,10$ (blue). Partial waves are nonzero only for even $\ell$, but we have analytically continued here for all $\ell$.
}
\label{fig:partialwaves}
\end{figure}

\noindent {\bf Large-spin behavior.}---The spin-dependence of partial waves in both our graviton amplitude ${\cal M}$ in \Eq{eq:MM} and the nongravitational theory $\overline {\cal M}$ in \Eq{eq:A1} is dictated by the combination $F_\ell^{(\alpha)} I_\ell^{(\alpha)}\!(M^2)$. At large even $\ell$, the asymptotic hypergeometric expansion yields, for $v\,{>}\,1$,
\be 
F_\ell^{(\alpha)}I_\ell^{(\alpha)}(v) \sim \frac{2^\alpha \Gamma(\alpha) (v^2\,{-}\,1)^{\frac{\alpha{-}1}{2}}}{v\,\ell^{\alpha{-}1} 
\,e^{(\alpha{+}\ell){\rm arcosh}\,v}}.
\ee
The nongravitational theory thus has partial waves~\eqref{eq:alR} that decay exponentially at large spin.
In contrast, for the gravitational theory the $F_\ell^{(\alpha)} I_\ell^{(\alpha)}(M^2)$ term enters Eq.~\eqref{eq:AA} with negative sign, while $I_\ell^{(\alpha)}(1) \,{=}\,\sqrt{\pi}\,\Gamma(\alpha{-}\frac{1}{2})/\Gamma(\alpha)$ provides the positive contribution.
Hence, at large $\ell$, the partial waves for our graviton amplitude simply scale like $F_\ell^{(\alpha)}$, which for $D\,{=}\,5$ is a constant and in higher $D$ scales as a power law ${\sim}\, 1/\ell^{2(\alpha-1)}$.
Thus, at least for $D\,{>}\,5$, our amplitudes generically satisfy the low-spin dominance observed in other known UV-complete theories~\cite{Bern:2021ppb,Arkani-Hamed:2020blm}, despite the presence of infinite towers of higher-spin states.
See Fig.~\ref{fig:partialwaves} for an illustration.
\newline

\noindent {\bf Discussion.}  In this paper, we have presented an infinite class of novel four-graviton amplitudes that satisfies the usual S-matrix unitarity requirements, including perturbative unitarity at high energies and positivity on the thresholds. It is straightforward to check that our four-point graviton amplitude in Eq.~\eqref{eq:MM} satisfies further low-energy constraints. For example, both the leading and subleading Weinberg soft theorems are obeyed, since the low-energy expansion takes the form~\footnote{The nongravitational amplitude in \Eq{eq:A1} has a very similar EFT expansion, but with the $\mathbb{R}^4$ stripped off and the indices $i,j,k$ starting at $1$ rather than $0$.}
\be 
\begin{aligned}
&{\cal M} = \kappa^2 \mathbb{R}^4 \sum_{i,j,k {=} 0}^\infty \mu_i \mu_j\mu_k s^{i-1} t^{j-1} u^{k-1} \\
&\! = \kappa^2 \mathbb{R}^4\! \left[\frac{1}{stu} + (\mu_1^2{-}2\mu_2) \left(\frac{1}{s}{+}\frac{1}{t}{+}\frac{1}{u}\right){+}\cdots \right]\!,
\end{aligned}
\ee
where $\mu_k = \sum_n g_n^2/m_n^{2k}$.
The term with the single poles only affects the sub-subleading soft theorem. 
Moreover, the Shapiro time delay can be computed as $\partial \chi/\partial E$, where $\chi$ is the phase of the S-matrix in the eikonal limit $s\,{\gg}\,t\,{=}\,{-}\vec q^2$, where $\vec q$ is Fourier-dual to the impact parameter~\cite{Camanho:2014apa,tHooft:1987vrq,Kabat:1992tb}. For pure gravity, $\left.{\cal M}_{\rm grav}\right|_{\rm eik} \propto +E^4/\vec q^2$, and our ${\cal M}$ in Eq.~\eqref{eq:MM} gives the same (causal) sign; moreover, when $\sum_n g_n^2 {=} 1$, as required for perturbative unitarity, we have $\left. {\cal M}\right|_{\rm eik} \propto E^0/\vec q^2$, so the leading-order time delay vanishes.

This paper leaves many compelling avenues for future investigation. Certainly, the existence of these amplitudes sharpens the necessity of addressing the elephant in the room: Are there some grounds on which one can rule out accumulation points---i.e., an infinite number of states with the same energy---in the S-matrix? If not, is it possible to construct a consistent theory from the amplitudes in Eq.~\eqref{eq:MM}? Note that our analysis indicates that accumulation points  in gravitational amplitudes are necessarily accompanied by an infinite tower of massless higher-spin states. 
These extra terms in the $t$ expansion of the $s\,{=}\,0$ residue are suppressed by the mass of the UV state, and therefore become relevant precisely when the entire massive accumulation-point tower does as well; it is therefore conceivable that these states are not separate entities from the full accumulation-point multiplet.
At face value, the presence of these terms does not violate no-go theorems \`a la Weinberg, since the higher-spin states are not asymptotic states in these amplitudes. This suggests that the states appearing at the accumulation points should not be considered as isolated asymptotic states in the corresponding theory either. This is suggestive of tensionless limits of string theory. In fact, the amplitude presented here is reminiscent of the twisted string amplitudes introduced in Refs.~\cite{Siegel:2015axg, Huang:2016bdd} (see also Refs.~\cite{Hohm:2013jaa,Bandos:2014lja,Gamboa:1989px,Gamboa:1989zc,Lee:2017utr,LipinskiJusinskas:2019cej}), where the right- and left-handed worldsheet modes have opposite signs in the two-point OPE, i.e.,
\begin{equation}\label{eq:Twist}
X(z_i,\!\bar{z}_i)X(z_j, \!\bar{z}_j){\sim} \log |z_{ij}|^2 {\rightarrow} \log z_{ij}  {-} \log \bar{z}_{ij}.
\end{equation}
This modification corresponds to a change in the boundary condition for the two-point function, which was later interpreted as a new vacuum on which the theory is quantized~\cite{Casali:2016atr}. As a result, e.g., the heterotic Yang-Mills amplitude becomes 
\be 
\pi\mathbb{F}^4\frac{1}{s(1\,{-}\,s)}\frac{1}{t(1\,{-}\,t)}(u\,{+}\,2)(u\,{+}\,1),
\ee
where $\mathbb{F}^4$ is a quartic form in the field strengths analogous to $\mathbb{R}^4$.
Note that $1/[s(1-s)]=1/s + 1/(1-s)$, which is exactly our $\mathcal{A}(s)$ in Eq.~\eqref{eq:Agrav} with $m^2\,{=}\,1$. Due to the factor $(u\,{+}\,1)(u\,{+}\,2)$, this amplitude does not have an accumulation point. It would be interesting to see if there is a variant of the twisted string amplitude that indeed gives our result, perhaps related to the asymmetrically twisted string amplitude of Ref.~\cite{Jusinskas:2021bdj}.

Finally, we note that our amplitude differs from string theory in that our ${\cal A}(s)$ remains asymptotically bounded both in the physical and unphysical regions of $s$, while the Virasoro-Shapiro amplitude does not, by virtue of the gamma functions.
On the other hand, the leading Regge trajectory in our amplitude has infinite slope (i.e., an infinite number of spins at a single mass).
It would be interesting to see if these two theories can be viewed as limits of a broader family and to sculpt out the full space of consistent triple-product amplitudes UV-completing gravity, of which both string theory and our class of amplitudes constitute examples.
We leave such investigations to future work.

\vspace{3mm}

\begin{acknowledgments}
\noindent {\it Acknowledgments:}
We thank Nima Arkani-Hamed, Cliff Cheung, Juan Maldacena, and Mark Wise for useful discussions and comments. 
Y.-t.H. is supported by MoST grant 109-2112-M-002 -020 -MY3. 
G.N.R. is supported at the Kavli Institute for Theoretical Physics by the Simons Foundation (Grant~No.~216179) and the National Science Foundation (Grant~No.~NSF PHY-1748958) and at the University of California, Santa Barbara by the Fundamental Physics Fellowship.
\end{acknowledgments}

\vspace{-5mm}

\bibliographystyle{utphys-modified}
\bibliography{product_amps}

\end{document}